# Heisenberg's Limit Phase Sensitivity in The Presence of Decoherence Channels

**Farzam Nosraty**

Research Institute for Applied Physics and Astronomy, University of Tabriz, Tabriz, Iran

[*]Email: farzam_nosrati91@ms.tabrizu.ac.ir

**Abstract**

In the present study, time evolution of quantum Cramer-Rao bound of entangled N00N state, as phase sensitivity, is determined by the aid of quantum estimation theory in the presence decoherence channels. Also, the dynamic quantum process as decoherence approach is characterized by quantum fisher information flow and entanglement amount in order to distinguish between Markovian and Non-Markovian process. The comparison between quantum fisher information and quantum fisher information flow assists to comprehend the phase sensitivity evolution corresponding to Non-Markovian and Markovian process. Furthermore, as result of backflow of information from the environment to system, the phase sensitivity corresponding memory effect of environment are revived after complete decay and increase in the few times.

## I. Introduction

In order to identify or estimate unknown parameters, an inference from the observed data about the parameters or about the system to which they are attributed is needed. The most important goal of metrology is enhancing the accuracy of data acquisition and inference, and then improving the estimation quality. Also, the basic physics of the intended system might show some limitations (i.e. bounds) on the ultimate achievable accuracy (commonly described through a 'Cramer-Rao inequality' [1]) or even might make new possibilities to exploit [2, 3]. The phase sensitivity in the quantum optical system, in another words, the best accuracy in detecting phase is bounded by the quantum version of Cramer-Rao bound (QCRB) which only depends on quantum state. The phase sensitivity of classical uncorrelated probe states is bounded by the so-called *shot noise limit* $\delta\varphi = 1/\sqrt{n}$. It has been known that the entangled state can be utilized to enhance the estimation accuracy of the parameters [4-8]. Accordingly, N00N state is used to improve phase sensitivity, which this use of N00N state achieves the so-called *Heisenberg limited* $\delta\varphi = 1/n$ sensitivity [9-15].

For closed quantum systems, it has been shown that the estimation precision depends on the underlying dynamics because all realistic quantum mechanical systems are in interaction with their surroundings. This inevitable interaction between a system and its environment typically results in the loss of quantum features and coherence [16-18]. As a result, the quantum estimation theory is expanded to the open quantum systems as a mix state in order to describe the effect of decoherence channel on the accuracy of parameter (phase sensitivity) [19-24]. Apart from the estimation scenario, much effort was done to describe, analyze and classify the quantum evolution into Markovian or non-Markovian process in the open quantum system [25, 26]. Markovian and non-Markovian processes are often described as memoryless and memory effect of environment. In particular, several so-called non-Markovianity measures were proposed which are based on divisibility distinguishability of states [27-31], quantum entanglement [28], quantum Fisher information (QFI) flow [32], fidelity [33], mutual information [34-36], channel capacity [37], and geometry of the set of accessible states[38].

In the current study, the mathematic framework of quantum estimation theory based on QCRB strategy is reviewed in Section 2 and the evolution of open quantum system as decoherence channels is reviewed in section 3. Then, Section 4 is dealt with determining QCRB (or QFI) for entangled N00N states in the presence of diverse decoherence channels. To relate phase sensitivity to dynamic properties of system and environment as Markovian and non-Markovian processes, QFI flow was and evolution of entanglement amount are obtained in section 5. For the purpose of making the analysis meaningful, it is assumed that the estimation procedure completion can take less than the decay rate does. Finally, Section 6 is concerned with conclusion.

## II. Mathematic framework of quantum estimation theory

In the quantum estimation strategy which is trying to infer the parameter $\varphi$ is encoded in the quantum state $\rho(\varphi)$. General measurement - mathematically represented by the elements of *positive value operator measure* (POVMs) $M_x$ - is performed to determine the outcome statistics [2, 3]. In the QCRB scenario, the bound to accuracy for any quantum measurement which means independence from choosing POVMs is aimed at estimating the parameter as a phase defined by QFI

$$\delta\varphi \geq \frac{1}{\sqrt{MF}}, \qquad F = \text{Tr}(L^2(\varphi;t)\rho(\varphi;t)) \qquad (1)$$

where $M$ is the repeat time of independent measurements and $L_\rho(\varphi)$ is Hermitian operator defined for any quantum state $\rho(\varphi)$ by symmetric logarithmic derivative (SLD) equation $\partial_\varphi \rho(\varphi;t) = (\rho(\varphi;t)L_\rho(\varphi;t) + L_\rho(\varphi;t)\rho(\varphi;t))/2$. Using the complete eigenbasis $\rho = \sum_k p_n(\varphi;t)|\psi_n(\varphi;t)\rangle\langle\psi_n(\varphi;t)|$, the SLD operator can be described as [19-23]:

$$L_\rho(\varphi) = 2 \sum_{n,m} \frac{<\psi_n(\varphi)|\partial_\varphi \rho(\varphi)|\psi_m(\varphi)>}{p_m(\varphi) + p_n(\varphi)} |\psi_n(\varphi)><\psi_m(\varphi)| \quad (2)$$

### III. The decoherence channels

The open quantum processes described by the time-local master equation $\partial_t \rho(t) = \mathcal{K}(t)\rho(t)$ where $\mathcal{K}(t)$ is *Lindbladian* super-operator given with

$$\mathcal{K}(t)\rho(t) = -i[H, \rho(t)] + \sum_i \gamma_i(t)\left[A_i \rho(t) A_i^\dagger - \frac{1}{2}\{A_i^\dagger A_i, \rho(t)\}\right], \quad (3)$$

$H$ is Hamiltonian of the system, $\gamma_i(t)$ are the decay rates, and $A_i(t)$ are the Lindblad operators describing the type of noise affecting the system. If all $\gamma_i(t)$ and $A_i(t)$ are time-independent, the master equation becomes convectional Markovian process. In another case, if $\gamma_i(t)$ and $A_i(t)$ parameters are time-dependent and somehow $\gamma_i(t)$ can be temporary negative, the master equation may exploit the non-Markovian process [16, 17]. The dynamic of quantum system can be represented in the terms of trace-preserving single qubit quantum operations which can be defined as the operator bases set including $\mathbf{1}, \sigma_z, \sigma_x$ and $\sigma_y$ [18] as fallow:

$$A_t(\mathbf{1}) = \mathbf{1} + f(t)\sigma_z, \qquad A(\sigma_z) = h(t)\sigma_z, \quad (4)$$

$$A(\sigma_x \pm \sigma_y) = g(t)(\sigma_x \pm \sigma_y).$$

The quantum operation of system as Dephasing, Depolarization and Spontaneous emission are evaluated which described by the time-independent master equation. Parameters, in equation, for noisy channels are defined as fallowing way;

$$\text{Dephasing: } f(t) = 0, h(t) = 1, g(t) = e^{-\gamma_1 t}$$

$$\text{Depolarization: } f(t) = 0, h(t) = e^{-\frac{2\gamma_2}{3}t}, g(t) = e^{-\frac{2\gamma_1}{3}t} \qquad (5)$$

$$\text{Spontaneous emission: } f(t) = 1 - e^{-\gamma_2 t}, h(t) = e^{-\gamma_2 t}, g(t) = e^{-\gamma_1 t}$$

where $T_2 = 1/\gamma_2$ is longitudinal decay time, the process which involves exchange of energy, $T_1 = 1/\gamma_1$ is the transfer dephasing time phase randomization which those are fundamentally irreversible. Also complete positivity requires that $T_1 \leq 2T_2$. In the time-dependent master equation, two-level quantum system interacting with a zero temperature relaxation environment coupled to a reservoir consisting of harmonic oscillators. The cavity modes can be described as a Lorentzian spectral density $J(\omega) = 2\gamma_0 \lambda^2 / 2\pi((\omega_0 - \omega)^2 - \lambda^2)$ where the spectral width of the coupling $\lambda$ is related to the correlation time of the environment $\tau_B \approx \lambda^{-1}$ and the time scale $\tau_B \approx \gamma_0^{-1}$ denoting the exchange of the system state. The time-dependent decay rate is given by

$$\gamma(t) = \frac{2\gamma_0 \lambda \sinh\frac{dt}{2}}{d \cosh\frac{dt}{2} + \lambda \sinh\frac{dt}{2}} \qquad (6)$$

with $d = \sqrt{\lambda^2 - 2\gamma_0\lambda}$ [16]. Similarly to construct a quantum process, the parameters are chosen as

$$h(t) = e^{-\lambda t}\left(\cosh\frac{dt}{2} + \frac{\lambda}{d}\sinh\frac{dt}{2}\right)^2, \quad f(t) = 1 - h(t) \quad g(t) = \sqrt{1-f(t)}. \tag{7}$$

Also, The generalized amplitude damping channel which describes the relaxation of a quantum system when the surrounding environment at finite temperature define by fallowing equation.

$$f(t) = (1 - 2e^{i\omega t})(1 - e^{-\delta t}), \quad h(t) = e^{-\delta t}, \quad g(t) = e^{-\delta t/2} \tag{8}$$

where δ is longitudinal decay time and ω is real number. It assumed that the decoherence channels are independent and identical from one probe qubit to the next, are continuously differentiable and time stationary, and commute with rotations about the σ axis.

## IV. Phase sensitivity in the presence of decoherence

In the previous section, the mathematic framework of quantum estimation strategy based on QCRB was reviewed. In this Section, the phase sensitivity based on QFI in presence of decoherence channels is evaluated. Decoherence channels are applied to the entangled N00N state, $|\psi> = \frac{1}{\sqrt{2}}(|N>_A |0>_B + |0>_A |N>_B)$, which N00N state is pure in the initial state. By Applying decoherence maps, the density operator is described as following:

$$\rho(t) = \frac{1}{2^{n+1}} \bigotimes_{j}^{n} \left(\mathbf{1}_j + (f(t) + h(t))\sigma_{z:j}\right) +$$

$$\frac{1}{2^{n+1}} \bigotimes_{j}^{n} \left(\mathbf{1}_j + (f(t) - h(t))\sigma_{z:j}\right) + \tag{9}$$

$$\frac{1}{2^{n+1}} \bigotimes_{j}^{n} (g(t))^n e^{in\varphi} (\sigma_{x:j} + i\sigma_{y:j}) +$$

$$\frac{1}{2^{n+1}} \bigotimes_{j}^{n} (g(t))^n e^{-in\varphi} (\sigma_{x:j} - i\sigma_{y:j}).$$

Also, the relative phase shift $e^{in\varphi}$ is accumulated when each photon in mode B in equation acquired a phase shift of $e^{i\varphi}$. By diagonalization density operator, the SLD operator in the equation (2) is obtained by the fallowing equation:

$$L = \frac{1}{2^n} \left(\frac{in}{\lambda_1(t) + \lambda_n(t)}\right) *$$

$$* \left( \bigotimes_{j=1}^{n} e^{in\varphi}(g(t))^n (\sigma_{x:j} + i\sigma_{y:j}) - \bigotimes_{j=1}^{n} e^{-in\varphi}(g(t))^n (\sigma_{x:j} - i\sigma_{y:j}) \right) \tag{10}$$

Thus, the upper bound of QCRB is obtained for each decoherence channels as fallowing equation

$$F = \eta N^2, \eta = (g(t))^n (a_{11} + a_{nn})/(\lambda_1(t) + \lambda_n(t)) \tag{11}$$

where the QCRB are determined by the eigenvalues $\lambda(t) = (\lambda_1, \lambda_2, \ldots, \lambda_n)$ and elements of density operator $a_{ii}$. The Fig. 1 shows *Heisenberg limit* phase sensitivity evolution in the presence of noisy channels, dephasing, depolarization and amplitude damping, in the case of time-independent decay rate $\gamma_0$ as Markovian process described by Equation. Consequently, phase sensitivity which is *Heisenberg's limit* in the initial state is lost after decay time and phase information is no longer accessible.

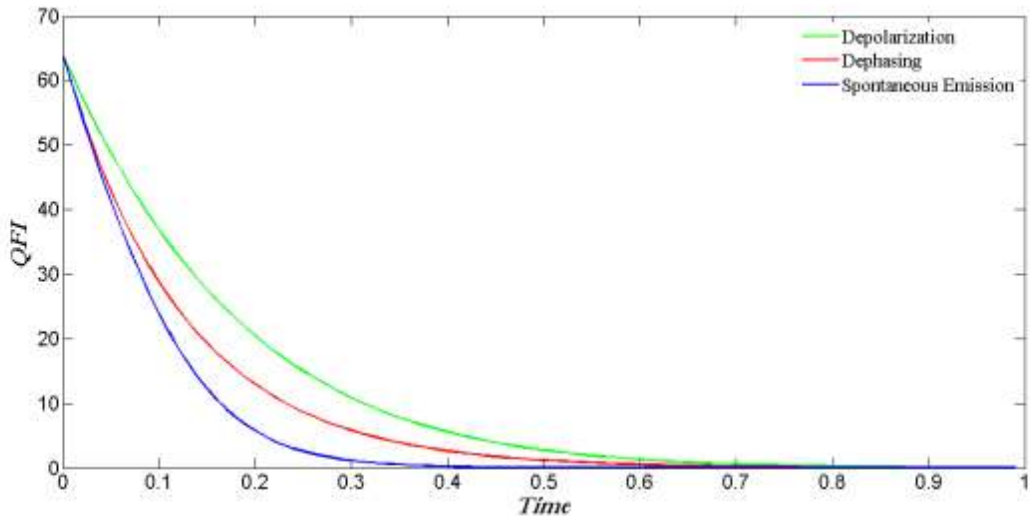

Fig. 1: QFI against $Time = \delta_1 t$ is plotted for eight qubits ($n = 8$) in presence of decoherence channels.

The QFI for time-dependent decay rate, reservoir consisting of harmonic oscillators, are plotted in Fig. 2 and Fig. 3. The evolution of phase sensitivity depend on underlying of dynamic of quantum system. The Fig. 2 is shown QFI in the weak coupling regime ($\lambda/\gamma_0 = 3$), which describe the phase sensitivity is decayed. The

Fig. 3 indicate the phase sensitivity in strong coupling regime ($\lambda/\gamma_0 = 0.1$) which describe the phase sensitivity is revived in the several times after complete decline and finally, the phase information is no longer accessible. Also, The QFI in the presence of generalize amplitude damping is shown in Fig. 4 ($\omega = 0.1$) and Fig. 5 ($\omega = 10$). The phenomenology parameters $\omega$ is responsible for underlying dynamics of open quantum system and oscillating phase sensitivity evolutions. In the other words, by increasing the $\omega$ parameter the phase sensitivity is rising in the certain times before complete decline.

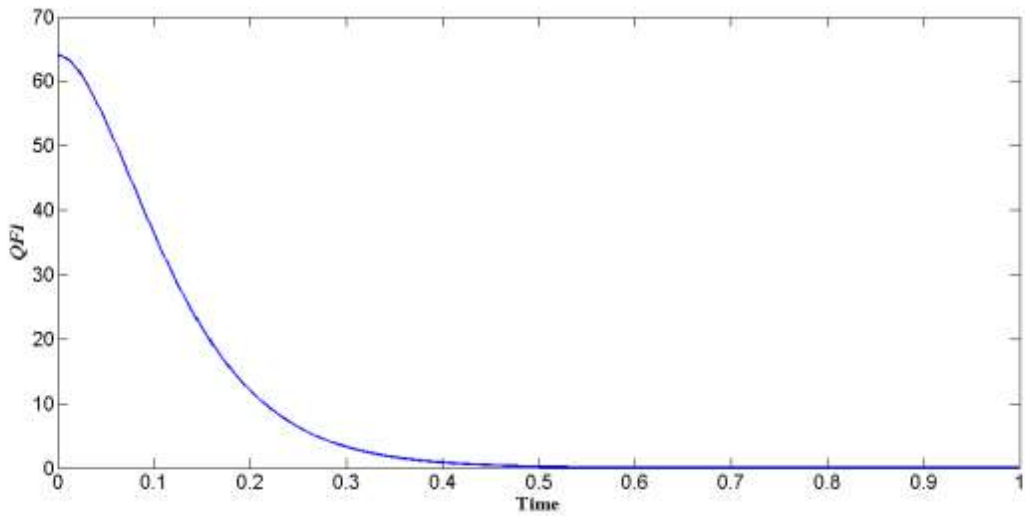

Fig. 2: QFI against $Time = \delta_1 t$ is plotted for eight qubits ($n = 8$) in presence of reservoir in the weak coupling regime ($\lambda/\gamma_0 = 3$)

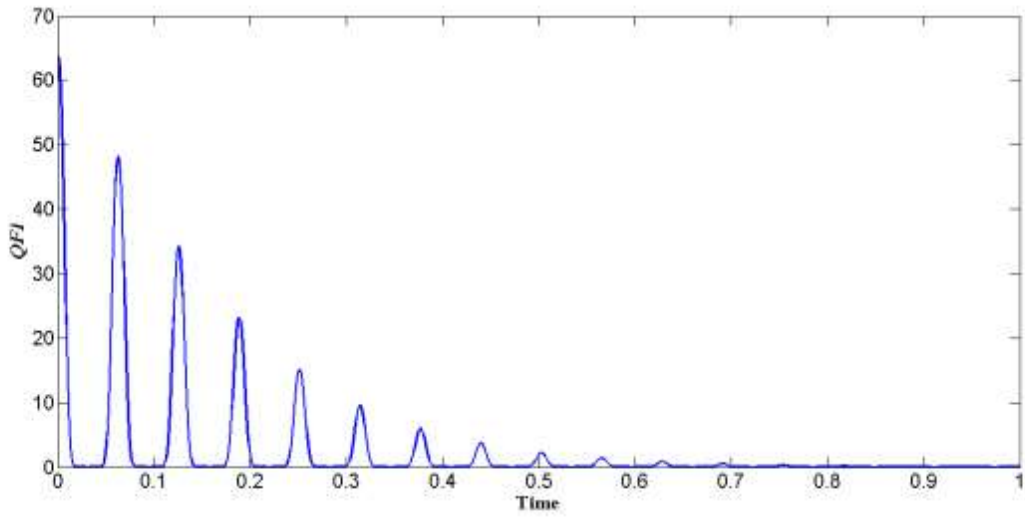

Fig. 3: QFI against $Time = \delta_1 t$ is plotted for eight qubits ($n = 8$) in the presence of reservoir in the strong coupling regime ($\lambda/\gamma_0 = 3$)

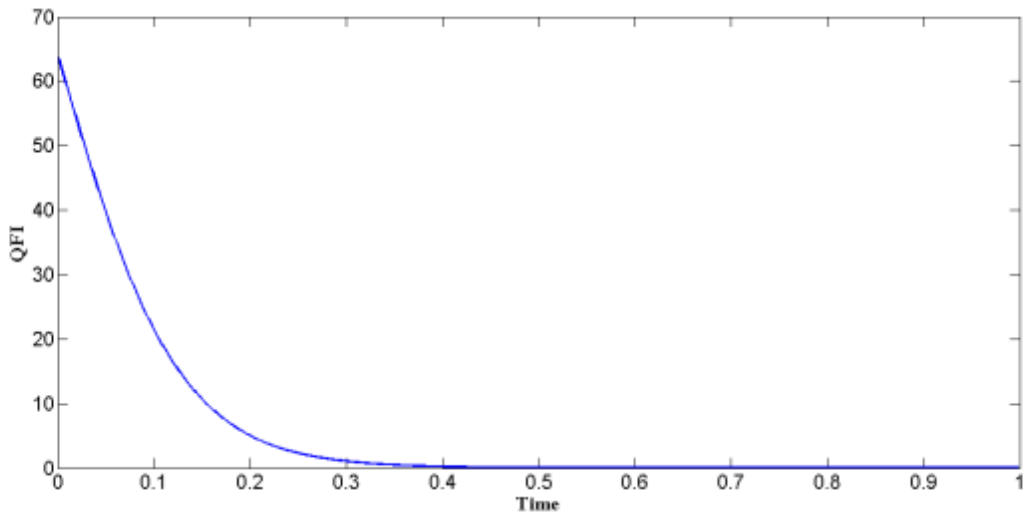

Fig. 4: QFI against $Time = \delta_1 t$ is plotted for eight qubits ($n = 8$) in the presence of generalized amplitude damping ($\omega = 0.1$)

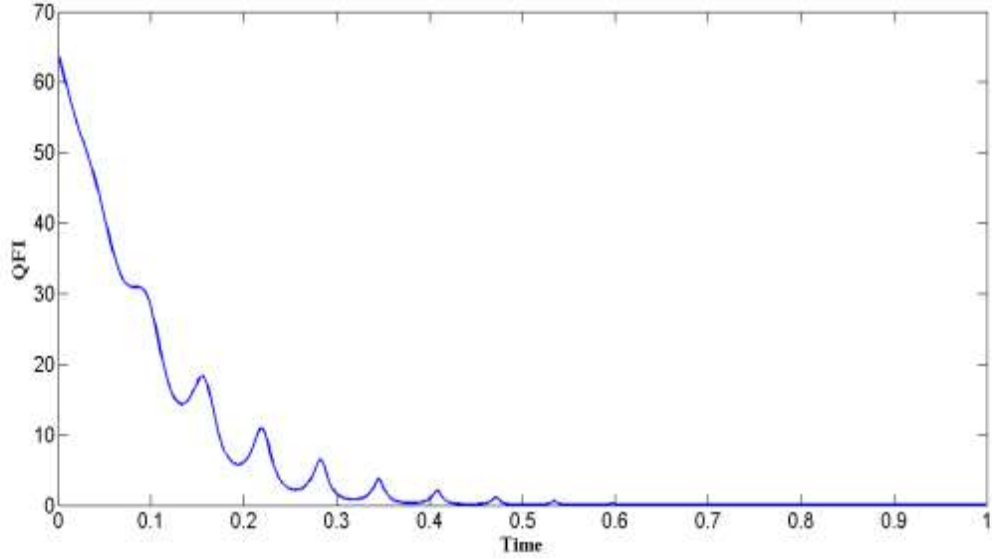

Fig. 5: QFI against $Time = \delta_1 t$ is plotted for eight qubits ($n = 8$) in the presence of generalized amplitude damping ($\omega = 10$).

## V. Markovian and non-Markovian processes

In this section, the QFI flow and entanglement amount are studied in order to relate the evolution phase sensitivity to dynamic of open quantum system, which correspond to Markovian and non-Markovian processes. In the parameter estimation approach, the (non)Markovianity of the open quantum system is characterized by introducing the QFI flow which is defined as the change rate of the QFI.

$$I := \partial F / \partial t \tag{12}$$

The QFI flow shows the evolution of information in the open quantum systems which it's exchange between systems and environment. Adopting the differential of Equation (1), the QFI flow is obtained as the fallowing equation:

$$I = \text{Tr}(\mathcal{L}(\partial_t \rho)) \tag{13}$$

where operator $\mathcal{L} := L(2\partial_\varphi - L)$ is defined. Sub-flow QFI information can be defined as explicit decomposition $I_i = \gamma_i(t) \mathcal{J}_i$ in which there is a direct relation between decay rate $\gamma_i(t)$ and QFI flow information. The sub-flow information is defined with

$$\mathcal{J}_i := -\text{Tr}(\rho[L, A_i]^\dagger \{L, A_i\}) \leq 0. \tag{14}$$

Non-positive QFI $I < 0$ indicating that all $\delta(t)$ is positive expresses that there is no backward information to open quantum system as a time-dependent Markovian process. If the total QFI flow $I(t)$ is positive at time $t$, it signifies that at least one of $\delta_i(t)$ is negative. In such cases, the QFI flows back to the open quantum system and consequently the non-Markovian behavior emerges [32]. Also, The characterization and identification of open quantum system can be comprehend by the evolution of correlation of quantum state as entanglement amount which leads to distinguish between Markovian and non-Markovian process. The measure of Markovianity can be exploiting the specific behavior of quantum correlations when a part of a composite system is subject to a local interaction that can be modeled as a trace-preserving map. The measure is proposed, that denoted by $I^{(E)}$, to quantifies the deviation from Markovianity in the evolution of the system. The expression is defined as fallowing Equation

$$I^{(E)} := \Delta E + \int_{t_0}^{t_1} \left|\frac{dE(\rho)}{dt}\right| dt$$

where $\Delta E = E(t) - E(t_0)$. The evolution of quantum system is Markovian if $I^{(E)} = 0$ otherwise is non-Markovian [28]. The Wooster's Concurrence is utilized in order to measure the entanglement amount of bipartite entangled states which varies from C = 0 for a separable state to C = 1 for a maximally entangled state [39].

The QFI flow are plotted respectively in the Fig. 6 and Fig. 7 in the weak ($\lambda/\gamma_0 = 3$) and strong ($\lambda/\gamma_0 = 0.1$) coupling regime. Regarding Equations (12), it can be stated that one reason for this Markovianity is the fact that the decay rate is positive at any time in the weak coupling regime. However, the non-Markovianity behavior shows up in the strong coupling regime, which because the time decay rate $\gamma_i(t)$ becomes negative in certain times. Also, the QFI flow are plotted in the presence of general amplitude damping for the both amount of $\omega = 0.1$ and $\omega = 10$. The Fig. 8 indicates the time-dependent Markovian process and by increasing the phenomenological parameter the non-Markovianity is raised up in the Fig. 9. In the both case, the backward information are comes from the environment $I > 0$ as memory effect and as result the phase sensitivities is revived after complete decay in Fig. 3 and increase in certain times in Fig. 5

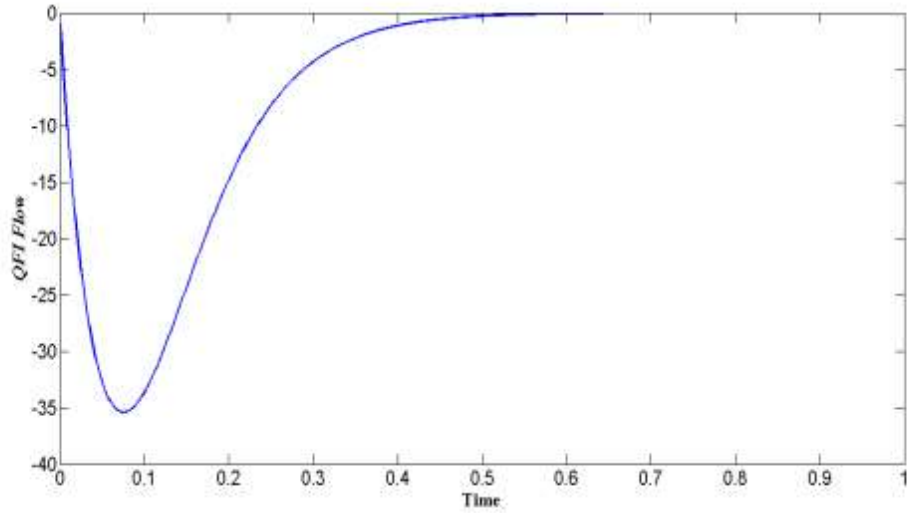

Fig. 6: QFI flow against Time $= \delta_1 t$ is plotted for eight qubits (n = 8) in the presence of reservoir consisting of harmonic oscillators in the weak coupling regime ($\lambda/\gamma_0 = 3$).

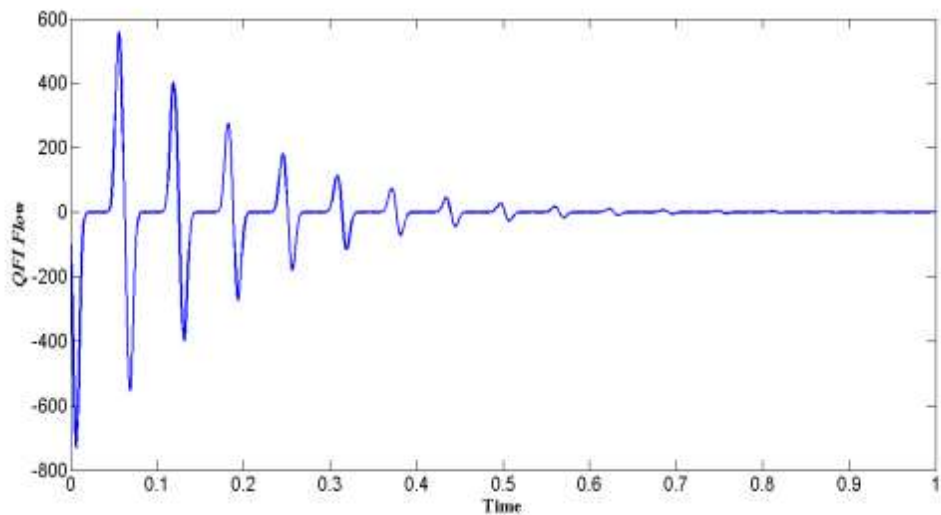

Fig. 7: QFI flow against Time $= \delta_1 t$ is plotted for eight qubits (n = 8) in the presence of reservoir consisting of harmonic oscillators in the strong coupling regime ($\lambda/\gamma_0 = 0.1$).

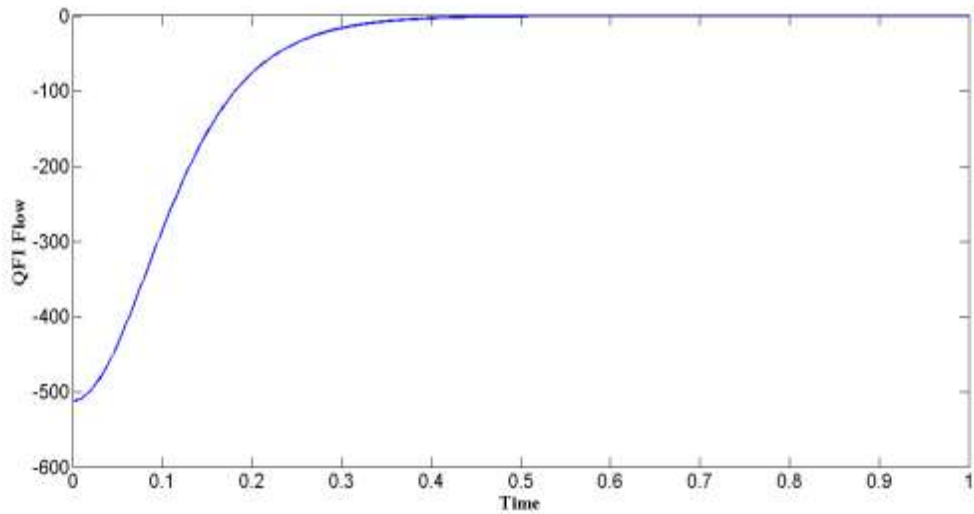

Fig. 8: QFI flow against Time = $\delta_1 t$ is plotted for eight qubits (n = 8) in presence of generalized amplitude damping ($\omega = 0.1$).

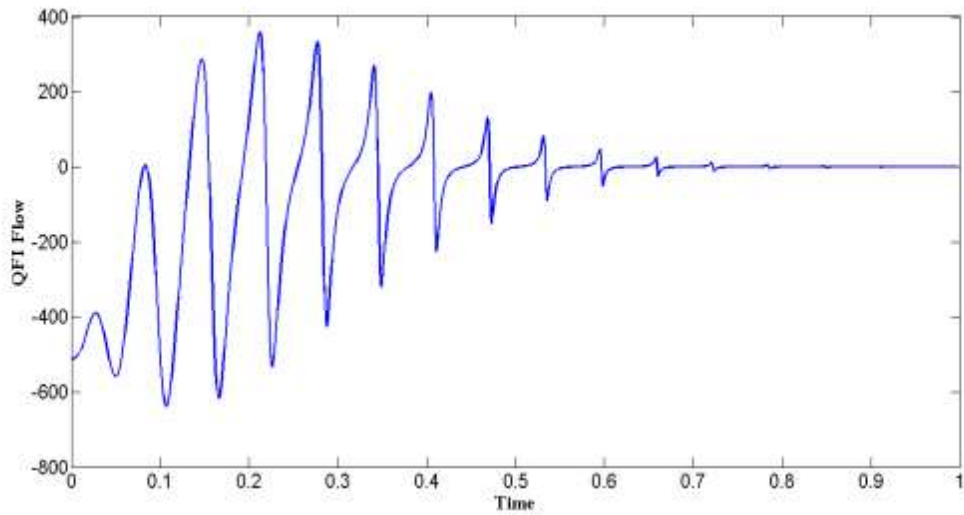

Fig. 9: QFI flow against Time = $\delta_1 t$ is plotted for eight qubits (n = 8) in presence of generalized amplitude damping ($\omega = 10$).

Besides, since the CP channels do not increase the amount of entanglement, the entanglement will be monotonically decreasing for Markovian evolutions, which shown in Fig. 10 and Fig. 11 ($\omega = 0.1$). However, the non-Markovianity behavior shows up in the strong coupling regime and by increasing phenomenological parameter. Also, if the evolution is non-Markovian, the requirement of strict monotonicity does no longer hold, the bipartite entanglement between photons to be revived for several times, Fig. 12. and be increased and decreased as a function of time Fig. 13. Besides, the Fig. 14 and Fig. 15, which indicate as measure of Markovianity $I(t)$, shows the situations that non-Markovianity rise.

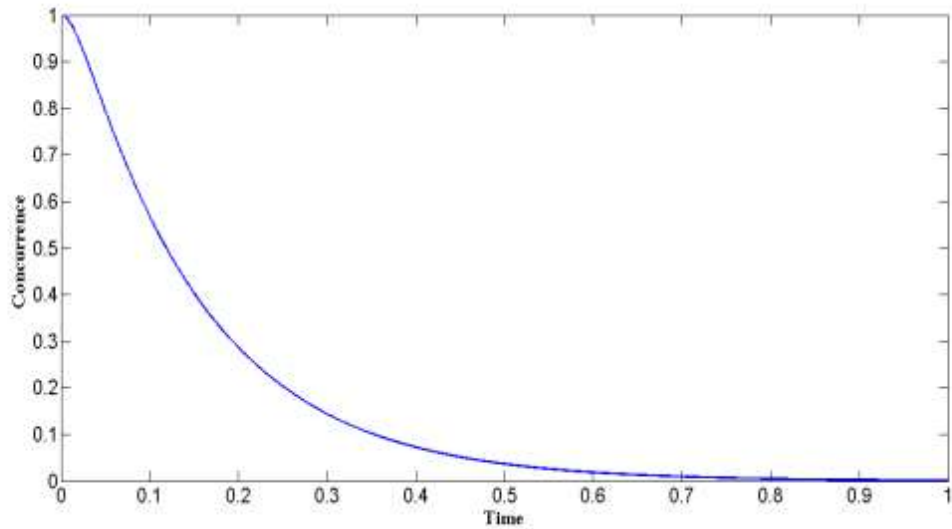

Fig. 10: Concurrence against Time $= \delta_1 t$ is plotted for two photons (n = 2) in the presence of reservoir consisting of harmonic oscillators in the weak coupling regime ($\lambda/\gamma_0 = 3$)

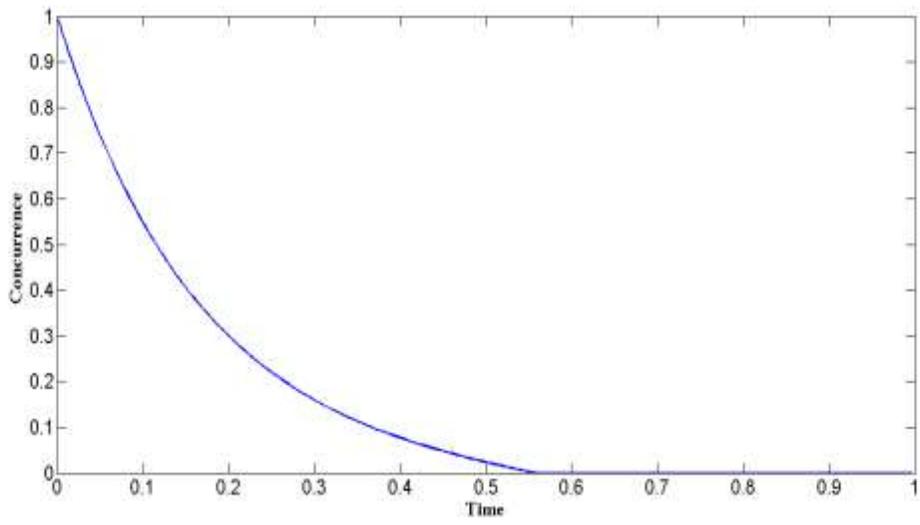

Fig. 11: Concurrence against Time = $\delta_1 t$ is plotted for two photons (n = 2) in presence of generalized amplitude damping ($\omega = 0.1$)

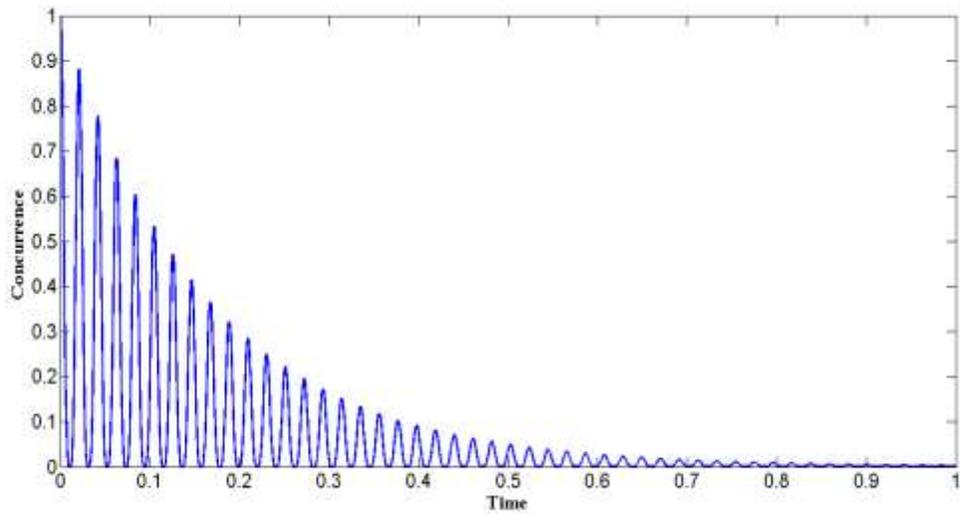

Fig. 12: Concurrence against Time = $\delta_1 t$ is plotted for two photons (n = 2) in the reservoir consisting of harmonic oscillators in the strong coupling regime ($\lambda/\gamma_0 = 0.1$)

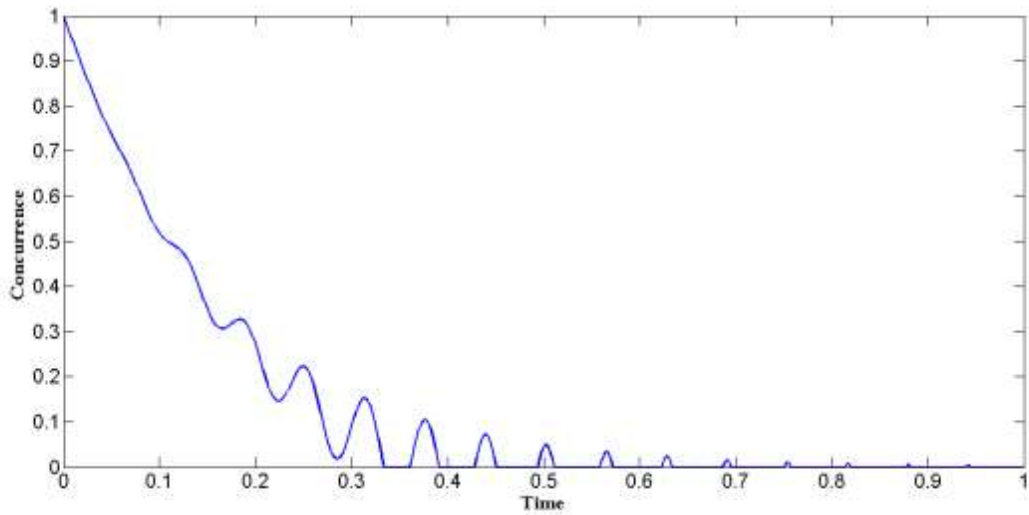

Fig. 13: Concurrence against Time = $\delta_1 t$ is plotted for two photons (n = 2) in presence of generalized amplitude damping ($\omega = 10$)

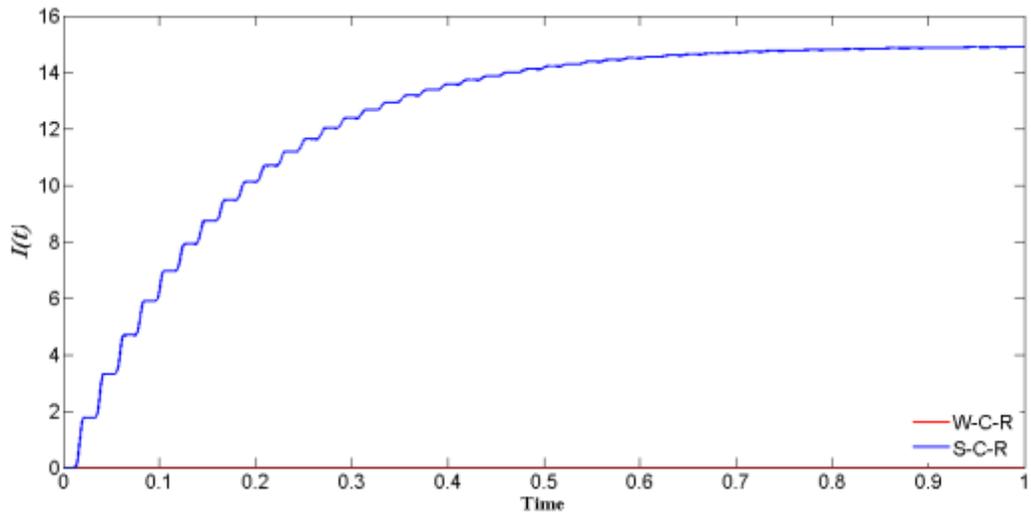

Fig. 14: $I(t)$ against Time $= \delta_1 t$ is plotted for two photons (n = 2) in presence of reservoir consisting of harmonic oscillators in the weak (red line) and strong coupling regime (blue line)

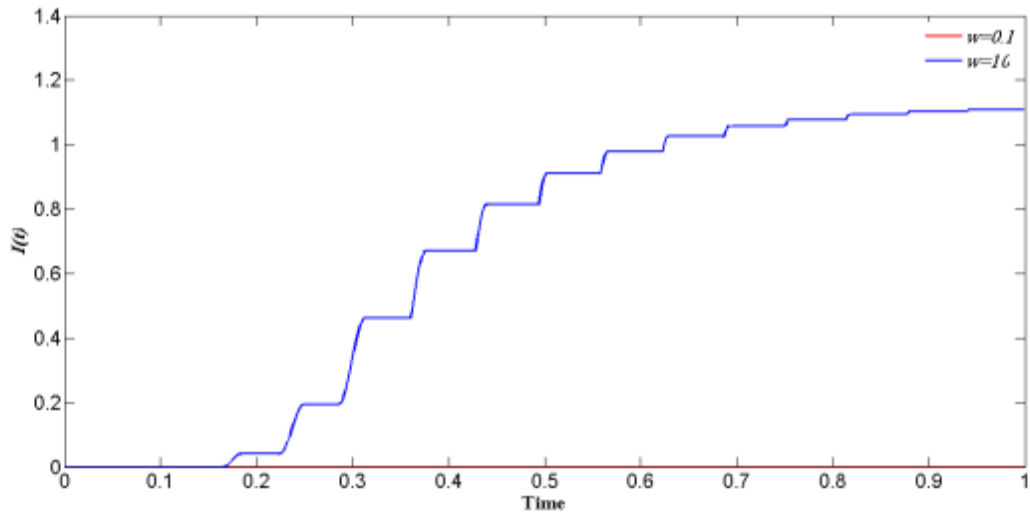

Fig. 15: $I(t)$ against Time $= \delta_1 t$ is plotted for two photons (n = 2) in presence of generalized amplitude damping ($\omega = 0.1$) (red line) and ($\omega = 10$) (blue line)

It can be asserted that the Markovianity depends on coupling constant of system and environment. Having found which process is Markovian and which one is not, it can be determined which phase sensitivity precisions belongs to the Markovian process and which one to the non-Markovian one. Therefore, the phase sensitivity as QFI is corresponded to non-Markovian process is more accessible and is maintained longer in the system than Markovian one.

## VI. Conclusion

In the present study, the time evolution of Heisenberg limit phase sensitivity was determined by the aid of estimation strategy in the presence decoherence channels. Besides, the Markovian and non-Markovian process was characterized by the quantum fisher information which shows the information exchange between system and environment. Also, the correlation between the photons in N00N state as entanglement amount is determined in the presence decoherence channels which leads to distinguish between Markovian and non-Markovian process. The comparison between quantum Fisher information and Non-Markovian measures led to the comprehension of the phase sensitivity corresponding to Markovian and non-Markovian processes. As a result, the Non-Markovian process phase sensitivity was revived in few times and more accessible and was maintained longer than the Markovian one.